\documentclass{aastex}          
\usepackage{spr-astr-addons}    
\usepackage{natbib}

\newcommand{\unit}[1]{\,{\rm #1 }}

\newcommand{\labequn}[1]{\label{eq:#1}}

\newcommand{\labsecn}[1]{\label{sec:#1}}

\newcommand{\equn}[1]{eq.(\ref{eq:#1})}
\newcommand{\equns}[1]{eqs.(\ref{eq:#1})}
\newcommand{\dequn}[2]{equs.(\ref{eq:#1})~and~(\ref{eq:#2})}

\begin{document}
%
\title{The Genesis of the Big-Bang and Inflation}

\shorttitle{Big-Bang and Inflation}
\shortauthors{R. K. Thakur}

\author{R. K. Thakur\altaffilmark{}} 
\affil{Retd. Professor of Physics \\
School of Studies in Physics and Astrophysics \\
Pt. Ravishankar Shukla University \\
Raipur - 492010, India}
\email{rkthakur0516@yahoo.com}


\begin{abstract}
The standard model of cosmology posits that some time in the remote past,
labelled as $t=0$, a Big-Bang occurred. However, it does not tell what caused 
the Big-Bang and subsequently the Inflation. In the present work the cause of the 
Big-Bang and Inflation is suggested on the basis of the hints provided by the
experimental findings at CERN and RHIC. The model used is singularity free
Newtonian, i.e., non-relativistic, oscillatory model of the universe in which
the \textit{space does not expand} whereas all the relativistic  cosmological 
models of the universe including the standard model, except the now discredited 
Einstein's static model, imply that apart from the matter and the radiation in
the universe the \textit{space} is also expanding. However, there is no observational 
evidence whatsoever of the expansion of the \textit{space} and as such, in all 
probability, the \textit{space} is not at all expanding. A critique of the singularity
theorems is also given on the basis of the experimental findings at CERN and RHIC
and it is emphasized that no gravitationally collapsing object can collapse to a
singularity, if it does, the time honoured  Pauli's exclusion principle would be
violated.   
\end{abstract}

\section{Introduction}
\labsecn{intro}
Gamow's Hot Bing-Bang (HBB) model of the universe has withstood the test of the 
time. It successfully explains the observed Hubble expansion of the universe, 
i.e., the recession of galaxies away from each other, the existence of 
the $2.73\unit{K}$ cosmic microwave background radiation (CMBR) as a relict of 
the big-bang, first predicted by \citet{Gamow1946,Gamow1948} and his collaborators
\citep{Alpher1948b,Alpher1948a,Alpher1948,Alpher1950} and subsequently observed 
by \citet{Penzias1965} and others, and the cosmic abundance of light elements
$H$, $D$, $^{3}He$, $^{4}He$, and $Li$ 
\citep{Steigman1976,Steigman1979,Olive1981,Yang1984,Boesgaard1985,
Audouze1987a,Audouze1987b}. Thus there are compelling evidences of the occurrence
of the big-bang in the remote past despite the desperate attempts of its detractors
to disprove it.

However, the so called standard cosmology based on the Friedmann models of the 
universe \citep{Friedmann1922,Friedmann1924} is plagued with a number of problems,
e.g., singularity \citep{Penrose1965,Hawking1970,Geroch1966,Geroch1968}, horizon
\citep{Rindler1956,Guth1981} and flatness\citep{Dicke1979,Guth1981}. Moreover, it 
does not tell what caused the big-bang. The cause of the big-bang remains a
mystery in the standard cosmology. Besides, it has no answer to the intriguing
question: What preceded the big-bang ?.

The implication of the singularity in the standard cosmology is that before the 
big-bang the universe was just a geometrical point of zero volume in the 3-space
of the four-dimensional space-time (i.e., in what is called \textit{space} in common
parlance), and the entire matter in the universe, if at all it was there before the 
big-bang, was squeezed at that point and occupied zero volume in the \textit{space}.
After the big-bang that point started expanding continually, i.e., its volume 
increased continually from its initial zero value to larger and larger values
and eventually reached the enormously large value of 
$\sim 15\times10^{84}\unit{cm^{3}}$ \citep{Allen1973} that the universe has today.
If at all there was no matter in the universe before the big-bang, then the matter
in the universe was also created by the big-bang. Essentially then, the implication
of the standard cosmology is that both matter and the \textit{space} were created
as a result of the big-bang.

Various attempts have been made to resolve the problems of singularity, horizon, and
flatness. \citet{Starobinsky1980}, and \citet{Israelit1989} have proposed singularity
free cosmological models. To resolve the difficulties of the horizon and flatness
associated with the standard model of cosmology inflationary models of the universe
have been invoked \citep{Guth1981,Linde1982,Linde1983,Albrecht1982}. Another attempt
at resolving the problems of singularity, horizon, and flatness is through the 
conformal transformation \citep{Infeld1945,Beem1976,Kembhavi1978,Narlikar1986}. In
1992 the author also proposed a singularity free model of the universe that not
only resolves the problems of singularity, horizon, and flatness but also attempted,
for the first time, to suggest the cause of the big-bang \citep{Thakur1992,Thakur1995}.

All the above cosmological models of the universe are based on one or another of the
relativistic cosmological models of the universe. These relativistic cosmological models 
of the universe are based on Einstein's field equations of the general theory of 
relativity (GTR) and hence assume that the geometry of the space-time of the universe is not
Euclidean, it is Riemannian. However, there is no sound physical justification for this
assumption. It may also be noted that all the relativistic cosmological models of the universe,
except the now discredited Einstein's static model of the universe, imply per se, that apart 
from the matter and the radiation in the universe the \textit{space} is also expanding. However,
there is no observational evidence whatsoever of the expansion of the \textit{space}.  The observed
recession of the galaxies away from each other and the expansion and consequent cooling
of the radiation produced during the big-bang as evinced by its relict, CMBR, have been
misconstrued as the evidences of the expansion of the \textit{space} also. Actually, they are 
only evidences of the expansion of the matter and the radiation respectively in the universe.
They are not at all the evidences of the expansion of the \textit{space}. This is obvious from 
the following analogy also. When a gas expands, the separation, i.e., the distance between
each and every pair of molecules in the gas increases. Thus, the increase in the distance
 between each and every pair of molecules in the gas implies only that the gas is expanding,
i.e., it is gradually occupying more and more space. It in no way implies that the \textit{space} in which the gas is embedded is also expanding with the gas. So, how can the gradual increase in
the distance between each and every pair of galaxies in the universe with the passage 
of time, i.e., the recession galaxies away from each other, be construed to be the evidence 
of the expansion of the \textit{space} in which they are embedded.? It certainly cannot be.

Actually, the purported expansion of the \textit{space} has been unwittingly assumed while
 postulating  that the space-time of the universe is Riemannian and as such the line element
$ds$, i.e., the ``distance'' between two neighboring points ($x^{0} + dx^{0},x^{1}+dx^{1},
x^{2} + dx^{2}, x^{3} + dx^{3}$) and ( $x^{0},x^{1},x^{2},x^{3}$) in the space-time is given by
the metric 
\begin{equation}
\labequn{ds}
 ds^{2} = g_{\mu\nu}dx^{\mu}dx^{\nu}
\end{equation}
where $\mu,\nu$ take on the values 0,1,2,3 and the summation over repeated indices is
implied. In \equn{ds} the  metric tensors $g_{\mu\nu}$ are functions of time co-ordinate
$x^{0} = ct$ also apart from that of the spatial co-ordinates $x^{1},x^{2},x^{3}$. This
implies that not only $ds$ but also the spatial separation $dl$ between each and every pair 
of spatial points in the universe is a function of time and consequently changes with time,
it cannot remain constant thereby implying expansion or contraction of the space. 
This is more obvious on perusal of the spatial parts of the 
metrics of various relativistic cosmological models of the universe \citep{Thakur2009}.
Consequently, no solution of Einstein's field equations
\begin{equation}
\labequn{rmu1}
R_{\mu\nu} - \frac{1}{2}Rg_{\mu\nu} = -\kappa T_{\mu\nu}
\end{equation}
can be static; every solution will be a function of time. It is surprising how this 
point escaped the genius of Einstein who was dismayed on obtaining non-static solution of
\equns{rmu1} when he used them to solve the ``cosmological problem''. At that time Einstein 
believed that the universe was static. Therefore, in order to obtain a static solution
Einstein modified his \equns{rmu1} to 
\begin{equation}
\labequn{rmu2}
R_{\mu\nu} - \frac{1}{2}Rg_{\mu\nu} + \lambda g_{\mu\nu} = -\kappa T_{\mu\nu}
\end{equation}
The term $\lambda g_{\mu\nu}$ in \equn{rmu2} is called `` cosmological term'' and
the constant $\lambda$ is called ``cosmological constant''; it has dimension $L^{-2}$.

That the expansion of \textit{space} is an inherent feature of Riemannian geometry is
further evident from de Sitter's solution of \equns{rmu2} for an empty space (i.e. for 
$T_{\mu\nu}=0$) which has nothing to do with the universe in which we live. This 
solution, known as the de Sitter Universe, is not static. The de Sitter universe
expands though it is empty! Obviously, the \textit{space} expands in the de Sitter 
universe, neither matter nor radiation, because it is empty. Two pertinent questions
in this connection are;
\begin{enumerate}
 \item What triggers off the expansion of the de Sitter universe, i.e., of the
empty \textit{space} ?
\item What maintains the continual expansion of the de Sitter universe, i.e., of the
empty \textit{space} ?
\end{enumerate}
The only possible answer to these questions is : The assumed Riemannian geometry of the space-time
! For, their is as yet no known physical process that can trigger off and maintain the
continual expansion of an empty \textit{space}. 

The de Sitter universe reveals another flaw in the conceptual foundations of the GTR. For,
according to the GTR, the curvature in space-time is produced by the gravitational field;
however, the space-time of the de Sitter universe is curved ( though the de Sitter universe
is spatially flat) despite the fact that there is no gravitational field in it since it is 
empty; existence of gravitational field requires presence of matter somewhere.

Even in the Friedmann models of the universe expansion of the empty universe, i.e., of the
\textit{space} is inherent. This is obvious from the following dynamical equation governing 
the Friedmann models of the universe \citep{weinberg1972}:
\begin{equation}
\labequn{rdot1}
\dot{R}^{2}(t) + k = \frac{8\pi G}{3}\rho(t)R^{2}{t}
\end{equation}
where $R(t)$ is tha scale factor of the universe (not the scalar curvature $R$ occurring
in \dequn{rmu1}{rmu2} above), $\rho$ the density of matter and radiation in the universe, 
and $k=0,+1 \mathrm{or},-1$ according as the space-time of the universe is spatially flat,
positively curved, or negatively curved respectively. From \equn{rdot1} we see that 
in the limiting case of $\rho(t)=0$ 
\begin{equation}
\labequn{rdot2}
\dot{R}^{2}= - k 
\end{equation}
This implies that even in the absence of matter and radiation $\dot{R}(t)\neq0$
when $k\neq0$. When $k=-1$, $\dot{R}^{2}(t)=1$ and hence $\dot{R}(t)=\pm 1$, 
which means the empty universe, i.e. the \textit{space} either expands or contracts.
However, for $k=+1$, the expansion of the empty universe is imaginary. For $k=0$,
$\dot{R}(t)=0$, i.e. the empty universe neither expands nor contracts. In other
words, when $k=0$, the \textit{space} neither expands nor contracts,  an
inference which is quite realistic since there is no observational evidence
of the expansion or contraction of the \textit{space}. This fact may be used as an 
argument in favour of the flat model of the universe by the relativistic cosmologists.

From the foregoing considerations it is evident that the purported expansion of the 
\textit{space} is a built-in feature of the relativistic cosmology and may be far from the 
reality, in all probability, the \textit{space} is not at all expanding. Consequently, none
of the relativistic cosmological models of the universe in general, and standard 
cosmology in particular, should be held sacrosanct. Moreover, it may be noted that the
problems of singularity, horizon, and flatness with which the relativistic cosmological
models, except the steady state model and its variants, are plagued stem from the assumption
that the geometry of the space-time of the universe is Riemnnian.

However, we are so much enamoured of the elegance of the Riemnnian geometry and the
mathematical formulation of the relativistic cosmology based on it that we are not
prepared to consider any alternative approach to cosmology. Attempts are being made
to geometrize even the whole of physics. `At one time it was even hoped that the rest
of physics could be brought into a geometric formulation, but this hope has met
with disappointment, and the geometric interpretation of the theory of gravitation
has dwindled to a mere analogy, which lingers in our language in terms like ``metric'',
``affine connection'', and ``curvature'', but is not otherwise very useful. The
important thing is to be able to make predictions about images on the astronomers' 
photographic plates, frequencies of spectral lines, and so on, and it simply doesn't 
matter whether we ascribe this predictions to the physical effect of gravitational
fields on the motion of planets and photons or to a curvature of space and time,
(The reader should be warned that these views are heterodox and would meet with 
objections from many general relativists.)'\citep{Weinberg1972a}. 

Of late, with a view to resolving some of the problems with which the standard 
cosmology is plagued, Cyclic models of the universe \citep{Steinhardt2001a,Steinhardt2001b,
Steinhardt2004,Frampton2006,Baum2007a,Baum2007b} and Loop quantum cosmology 
\citep{Bojowald2001,Bojowald2005,Bojowald2007,Ashtekar1998,Ashtekar2001,
Ashtekar2003,Ashtekar2007} have been ingeniously invented. However, they are based on
dubious assumptions. Moreover, so far none of the predictions of these models has been
observationally validated. But they certainly testify to the mathematical ingenuity of
their innovators. However, all these models have their roots in geometry rather than in 
physics. Some of these models are based on spaces of more than 4 dimensions.
But they do not give the physical significance of the higher dimensions, i.e., they 
do not spell out what physical entity these higher dimensions represent. Some are 
based on string theory which has not yet been validated experimentally. Despite their
mathematical elegance Cyclic models and Loop quantum cosmology appear to be quite far-fetched.

However, the cause of the big-bang and inflation cannot be explored through geometry, 
it can be explored only through physics. In this connection it may be noted that subtle
hints as to the origin of the big-bang and inflation are provided by the ongoing experiments
at CERN, the European Organization for Nuclear Research, at Geneva and at RHIC, the 
Relativistic Heavy Ion Collinder, at Brookhaven National Laboratory (BNL) in Upton, New York.
The cause of the big-bang and inflation suggested here is based on the hints provided by the
experimental findings at CERN and RHIC as well as on the well established laws of physics.
The model of the universe used is Newtonian, i.e., non-relativistic, oscillatory model
in which the \textit{space does not expand}.

\section{Fundamental Constituents of Matter}
Matter is composed of hadrons and leptons. Hadrons are either bosons or fermions having 
strong interactions as well as electroweak interactions whereas leptons are fermions which
 do not have strong interactions; they have only electroweak interactions. Leptons are 
point-like objects having radii not greater than $10^{-16} cm$ \citep{Barber1979}. However, hadrons
 have finite radius, e.g., the radius of a protons is about $10^{-13} cm$ \citep{Hofstadter1955}. 
Whereas leptons are fundamental particles, hadrons are not; they are composite particles; 
they are composed of quarks $(q)$ and anti-quarks $(\bar{q})$. Hadrons are classified into
 two categories, viz., mesons and baryons. Mesons are bosons having baryon number zero 
whereas baryons are fermions having baryon number different from zero. Quarks are fermions 
and have spin $\frac{1}{2}$; they occur in six flavours; viz.,
 $u$ (up, mass $m = 1.5 \mathrm{to} 3.0 MeV$, charge $q = \frac{2}{3} e$),
 $d$ (down, $m = 3$ to $7 MeV$, $q = -\frac {1}{3} e$ ), $s$ (strange, $m= 95 \pm 25 MeV$,
 $q = -\frac {1}{3} e$), $c$ (charm, $m= 1.25 \pm 0.09 GeV$, $q = \frac {2}{3} e$),
 $b$ (bottom or beauty, $m= 4.20 \pm 0.07 GeV$, $\overline{MS}$ mass, $4.70 \pm 0.07 GeV, IS$
 mass, $q=-\frac{1}{3} e$), $t$ (top or truth, $m= 174.2 \pm 3.3 GeV$, direct 
observation of top events; $172.3 \pm  ^{10.2}_{7.6} GeV$, standard model electroweak fit,
 $q=\frac{2}{3} e$) \citep{Balantekin2006}. Existence of quarks of all the six flavours has
 been established experimentally beyond any shade of doubt. Each quark $(q)$ has 
a corresponding anti-quark $(\bar{q})$. Moreover, each flavour of quarks occurs 
in three primary colours, red, green, and blue. It may be noted, however, 
that colours, red, green, and blue of quarks, and anti-red, anti-green, and 
anti-blue of anti-quarks  have nothing to do with the actual visual colours, they 
are merely labels for describing the additional internal degree of freedom of quarks.

Baryons are composed of three quarks $(qqq)$ and anti-baryons of three anti-quarks
 $(\bar{q}\bar{q}\bar{q})$ whereas mesons are composed of a quark $q$ and an anti-quark
 $\bar{q}'$ ($q$ and $q^{'}$ need not be of the same flavour). For example, a proton 
is composed of two up and one down quark $(p: uud)$, neutron of one up and two down 
quarks $(n: udd)$, pion $\pi^{+}$ of $u$ and $\bar{d}$ $(\pi^{+}:u\bar{d})$, 
and $\pi^{-}$ of $d$ and $\bar{u}$ $(\pi^{-} : d\bar{u})$.

Hadrons are colourless, i.e., white. White means all the three primary colours 
are equally mixed. This means each baryon contains quarks of all the three colours, 
whereas a mesons contains a quark of a given colour and an anti-quark of the corresponding
 anti-colour so that each combination is over all white.

\section{Quantum Chromodynamics}
The concept of colour first introduced by \citet{Greenberg1964} to account for the 
apparent violation  of the spin-statistics theorem in case 
of $\Delta^{++}$ and $\Omega^{-}$ resonances 
plays a fundamental role in accounting for the interaction between quarks. The remarkable 
success of quantum electrodynamics (QED) in explaining the interaction between electric 
charges to an extremely high degree of precision motivated physicists to explore a similar 
theory for strong interactions. The result is quantum chromodynamics (QCD), a non-Abelian 
gauge theory (Yang-Mills theory) which closely parallels QED.

Drawing analogy from electrodynamics, \citep{Nambu1966} postulated that the three quark colours 
are the charges (the Yang-Mills charges) responsible for the inter-quark force just as 
the electric charge is responsible for the electromagnetic force between charged particles.
 The analogue of the rule that like charges repel and unlike charges attract each other
 is the rule that like colours repel, and colour and anti-colour attract each other. 
Apart from this, there is another rule in QCD, which states that different colours 
attract if the quantum state is antisymmetric and repel if the quantum state is symmetric
 under exchange of quarks.

Since colours serve as the Yang-Mills charges, each quark flavour transforms as a 
triplet of $SU_{c}(3)$ group that causes transitions between quarks of the same flavour
 but different colours. However, the $SU_{c}(3)$ Yang-Mills theory requires the introduction
 of eight new spin 1 gauge bosons called gluons. Gluons carry colour and as such they 
strongly interact with each other. Moreover, gluons couple to the left-handed and 
right-handed quarks in the same manner since the strong interactions do not violate 
the law of conservation of parity.

Just as the electromagnetic force between electric charges arises due to the exchange of 
photons, a massless vector (spin 1) boson, the force between coloured quarks arises due 
to the exchange of gluons. However, there is a striking difference between QED and QCD.
 Whereas the force between electric charges decreases with increasing distance, the force 
between quarks increases with increasing distance. This phenomenon leads to two important 
features of QCD, viz., asymptotic freedom \citep{Gross1973a,Gross1973b,Politzer1973} and 
infrared slavery \citep{Alabiso1976,Chaichian1981}.
 Asymptotic freedom means that inter-quark force fades away as two quarks approach each 
other infinitely closely. Infrared slavery of quarks is responsible for the 
confinement of quarks in hadrons, and for their being elusive. It has been suggested that 
the attractive force between two quarks increases with increasing separation at the 
rate of $1 GeV$ per fermi.

According to QCD at extremely high temperature and/or density, the hadronic matter undergoes 
a phase transition; from the normal hadronic phase it goes over to the quark-gluon 
plasma (QGP) phase. This phase consists of (almost) free quarks and gluons. The transition 
temperature for this phase transition was first predicted by the lattice gauge theory 
of QCD to be $\sim 2 \times 10^{12} K; (1.90 \pm 0.02) \times 10^{12} K$ according to the 
more exact calculations. This transition temperature is approximately equal to $175 MeV$ 
corresponding to an energy density of a little less than $1 GeV/fm^{3}$.

\section{Findings at CERN and RHIC}
The notion of QGP is not just a figment of imagination. Efforts have been afoot at 
CERN and RHIC for quite some time to create QGP in the laboratory. Pioneering attempts 
to create QGP were first made at CERN's Super Proton Synchrotron (SSP) in 1980's and 1990's. 
In 1994 the lead beam programme was started at CERN. A beam of $33 TeV$ 
(equivalent to $160 GeV$ per nucleon) lead ions from the SSP was used in the programme.
 In the programme seven groups of scientists, viz., NA44, NA45/CERES, NA49, NA50,
 NA52/NEWMASS, WA97/NA57 and WA98 collaborated and measured different aspects of 
lead-lead and lead-gold collision events. A report released by CERN on Feb. 10, 2000 said 
that by smashing together lead ions at CERN's accelerator at temperatures $100,000$ times 
as hot as sun's centre ( i.e., at $T \sim 10^{12} K$), and energy densities never before 
reached in laboratory experiments, a team of $350$ scientists from institutes in $20$ 
countries succeeded in isolating tiny  components called quarks from more complex particles 
such as protons and neutrons. 

"A series of experiments using CERN's lead beam have presented compelling evidence for 
the existence of a new state of matter $20$ times denser than nuclear matter, in which 
quarks instead of being bound up into more complex particles such as protons and neutrons, 
are liberated to roam freely", the report said.

Presenting "Evidence for a new state of Matter : An Assessment of the Results from the 
CERN Lead Beam Programme" \citet{Heinz2000} said "A common assessment of the collected data 
lead us to conclude that we now have compelling evidence that a new state of matter has 
indeed been created, at energy densities which had never been reached over appreciable 
volumes in laboratory experiments before and which exceed by more than a factor $20$ that 
of normal nuclear matter. The new state of matter found in heavy ion collisions at the 
SSP features many of the characteristics of the theoretically predicted quark-gluon plasma".

When two large nuclei, e.g., that of lead or gold, are accelerated to ultrarelativistic 
speeds and slammed into each other, they largely pass through each other. However, after 
the collision a hot volume, called "fireball", is created. This fireball 'is in a state 
of tremendous explosion, with expansion velocities exceeding half the speed of light, 
and very close to local thermal equilibrium at a temperature of about $100-120 MeV$. 
This characteristic feature gave rise to the name "Little Bang". The observed explosion 
calls for a strong pressure in earlier collision stages' \citep{Heinz2000}.

The programme of creating QGP at RHIC began in summer of 2000. On June 18, 2003 a special 
scientific colloquium was held at the BNL to discuss the latest findings at RHIC. At the 
colloquium it was announced that in the detector system STAR (Solenoidal Tracker AT RHIC) 
head-on collision between two beams of gold nuclei of energies $130 GeV$ per nuclei resulted 
in the phenomenon called "jet quenching". STAR as well as three other detector experiments 
at RHIC, viz., PHENIX, BRAHM, and PHOBOS, detected suppression of "leading particles", highly 
energetic individual particles that emerge from the nuclear fireballs in gold-gold collisions. 
Jet quenching and leading particle suppression are signs of QGP formation. The findings of 
the STAR experiment were presented at the BNL colloquium by Berkeley Laboratory's NSD 
(Nuclear Science Division) physicist Peter Jacobs. It was also reported by \citet{Aronson2005} 
in April 2005. Subsequently, at the April 18, 2005 meeting of the American Physical 
Society at Tampa, Florida the four collorations, viz., STAR, PHENIX, PHOBOS, and BRAHMS 
presented the evaluation of their experimental findings, which were later published in 
Nuclear Physics $A757$ (STAR Collaboration : \citet{Adams2005}, PHENIX Collaboration : \citet{Adcox2005}; 
PHOBOS Collaboration : \citet{Back2005}; BRAHM Collaboration : \citet{Arsene2005}).

The new experiments, ALICE, ATLAS, and CMS running on CERN's Large Hadron Collider (LHC) are 
continuing the study of the properties of the QGP.

\section{The cause of the Big-Bang and Inflation} 
The standard cosmology posits that some times in the remote past, labelled as $t=0$, a gigantic 
explosion, the so called big-bang, occurred as a result of which the universe was heated to an 
enormously high temperature. However, it does not answer the intriguing question : What caused 
the big-bang? Or, what physical process heated the universe to an enormously high temperature?

Attempt is being made here to suggest the cause of the big-bang and the inflation on the basis 
of the experimental findings at CERN and RHIC 
within the framework of a singularity free Newtonian, i.e., the non-relativistic, oscillatory 
model of the universe in which the \textit{space does not expand}. According to the proposed model, 
prior to the big-bang the entire matter in the universe was gravitationally collapsing continually. 
Consequently, with the continual collapse of the matter in the universe gravitational energy was 
released continually which heated the matter and raised its temperature continually and thereby 
increased the energy of the particles comprising the matter in the universe continually. 
Moreover, with the continual collapse of the matter in the universe the density of the matter 
in the universe also increased continually with consequent rise in energy density of the matter 
in universe.

When the temperature of the matter in universe reached the transition temperature
 $T \sim 2 \times 10^{12} K$, which amounts to an energy of $\sim 175 MeV$ per particle 
and corresponds to an energy density of a little less than $1 GeV/fm^{3}$, the entire 
hadronic matter in the universe underwent a phase transition, from the hadronic  phase 
it passed over to the QGP phase, which consisted of the asymptotically free 
quarks : u,d,s,c,b,t and gluons. Thus, when the temperature of the collapsing matter 
in the universe $T \geq 2 \times 10^{12} K$, the entire matter in universe was in the form of QGP permeated by 
leptons i.e., it consisted of spin $\frac{1}{2}$ quarks u,d,s,c,b,t which interacted 
through the colour force generated by the gluons as well as through the electroweak force, 
and   spin $\frac{1}{2}$ leptons, $e,\mu,\tau$ and their neutrons 
$\nu_{e}, \nu_{\mu}, \nu_{\tau}$ which interacted through the electroweak force only. 
In this way, bulk of the gravitational energy 
released during the gravitational collapse of the matter in the universe was utilized in 
deconfinement of quarks from hadrons, i.e., in disintegrating the hadrons into quarks and 
gluons. In other words, the gravitational energy released during the collapse liberated 
the quarks from the \textit{infrared slavery} and delivered them \textit{asymptotic freedom}. As the matter 
collapsed further the additional gravitational energy released during the collapse was 
utilized in heating the QGP permeated by leptons, i.e., in energizing the quarks and leptons in 
the universe. Thus, the gravitational energy released during the collapse was locked in the QGP 
permeated by leptons.

However, the collapse of the universe to a singularity, i.e., to a point in the 3-space 
(i.e., in the \textit{space} ) was averted; it was inhibited by Pauli's exclusion principle. 
The universe could not collapse to a point otherwise all the fermions, viz., quarks and 
leptons, of each and every species (i.e., flavour) would have been crammed into that point which 
could be occupied, according to Pauli's exclusion principle, by at most only two fermions of any 
species, one with, say, spin "up" and the other with spin "down". In other words, had the universe 
collapsed to a singularity, Pauli's exclusion principle would have been violated. However, Pauli's 
exclusion principle is inviolable and as such the universe could not collapse to 
 a singularity. Besides, had the entire matter in the universe collapsed to a singularity,
 i.e., to a point in the $3-$space, the uncertainty in each component of each and every 
particle of the matter in the universe would have been zero. Consequently, according to Heisenberg's 
Uncertainty principle, the uncertainty in the corresponding component of the momentum of each and 
every particle in the universe would have been infinite. This would have resulted in each and every 
particle of universe having infinite momentum and infinite energy. This can also be seen by the fact 
that had the entire matter in the universe collapsed to a singularity, the inter-particle separation
 $s$  between each and every pair of particles in the universe would have been zero. As the de 
Broglie wavelength $\lambda$ of any particle in the universe would be less than or at most equal to 
$s$, $\lambda = \frac{h}{p} \le s $, where $h$ is Planck's constant and $p$ the magnitude of the 
momentum of the particle. Consequently, when 
$s \to 0, p \to \infty$ and with it the energy of the particle $E \to \infty$. As particles of 
infinite energy and momentum cannot remain frozen at a point, i.e., stay put at a point forever, 
the collapse of the entire matter in the universe to a singularity could not occur on this count also.

Incidently, it may be pointed out that the singularity theorems of Penrose, Hawing, and Geroch do not 
tell what happens eventually to a gravitationally collapsing massive object, e.g., the universe, or a
 black-hole, after it collapses to a singularity. However, it is implicit in their theorems, that after 
collapse to a singularity the object stays put at the singularity forever. But can it? The above 
considerations show that if at all a massive object collapses to a singularity, it cannot stay put 
there forever, otherwise there would not have been the big-bang and we would not have been around 
today. However, if the singularity theorems of Penrose, Hawking, and Geroch are valid, then the 
pertinent question is : What happens eventually to a massive object after it collapses to a 
singularity? Has the GTR any answer to this question?

The snag is that while arriving at the singularity theorems the general relativists had all 
along given cognizance to gravitational interaction only, they ignored the other interactions, 
especially the strong interaction, viz., the QCD, regarding them as negligible in comparison 
to the gravitational interaction. But the fact is that at ultra-high energies and ultra-high 
densities the QCD rules the roost, not the gravitation as is obvious from the fact that the 
coupling constant of the strong nuclear interaction is many orders of magnitude larger than 
that of the gravitational interaction. This is also obvious from the experimental findings at 
CERN and RHIC. Moreover, while arriving at the singularity theorems they all along treated the 
matter in the gravitationally collapsing object as a classical fluid, they completely ignored 
its microscopic structure and its quantum mechanical behaviour which have far reaching consequences 
at ultra-high energies and ultra-high densities. Furthermore, the GTR has been validated 
experimentally only in the weak field limit, it has not yet been validated experimentally 
in the domain of strong gravitational fields.

If the gravitationally collapsing matter in the universe could not eventually collapse to a 
singularity, then another intriguing question is : What happened to it in the final stages 
of the collapse? Hints as to the answer to this 
question is provided by the experimental findings at CERN and RHIC given in Section 4 above.
The answer is the following:

In the final stages of the collapse innumerable collisions between the ultra-high energy nuclei 
in the ultra-dense matter in the universe occurred and at each and every collision point a fireball 
was created after the collision which presumably contained the QGP. Thus in the final stages of 
the collapse innumerable fireballs were created. These fireball were in a state of tremendous 
explosion and consequently each and every one of them exploded with a "Little Bang". The 
cumulative effect of the innumerable "Little bangs" so produced was the "Big-Bang". And, after 
the explosion, each and every fireball expanded with speeds exceeding half the speed of light 
in vacuum resulting in the inflationary expansion of the matter in the universe, i.e., the 
inflation.

Subsequently, with the expansion of the fireballs inter-quark separation between quarks increased 
and with it the attractive force between them also increased so much so that eventually all the 
quarks lost their asymptotic freedom and were subjected to infrared slavery resulting in their 
enslavement, i.e., their confinement, in hadrons. In this way all the quarks in the QGP were 
hadronized. As a result of this hadronization of all the quarks in the universe tremendous amount 
of energy was released which was earlier locked in the QGP before the big-bang.

\section{Discussion}
\labsecn{discussion}
In the oscillatory model of the universe the big-bang is preceded by the contracting phase
of the matter in the universe, and after the big-bang the matter in the universe expands 
continually again, but its velocity of expansion is continually decelerated due to
the opposing gravitational force. Eventually, the expansion comes to a halt and thereafter
the matter in the universe starts contracting continually again due to the self-gravitation
up to a certain minimum volume to be followed by another big-bang and expansion. The sequence 
of contraction, big-bang, expansion, contraction, big-bang ----- is repeated ad infinitum. 
However, in the model proposed here singularity never occurs, the matter in the universe 
never collapses to a singularity. Moreover, the \textit{space}  neither expands nor contracts, only 
the matter in the universe undergoes oscillations with successive phases of contraction and 
expansion occurring perpetually.

The detractors of the oscillatory model of the universe have the objection that it would 
violate the second law of thermodynamics according to which entropy only increases, it
does not decrease, in any process and as such the entropy of the universe would build up
from oscillation to oscillation resulting, eventually, in the " heat death " of the 
universe. However, this objection is not valid. By definition, the universe is a closed 
system, there is nothing outside the universe. The cycle of contraction and expansion of 
the matter in the universe is an adiabatic process since heat neither enters the system
from outside nor leaves the system. Moreover, it is a reversible processes. Thus, the 
sequence of contraction, big-bang, expansion, contraction, big-bang ..... of the 
matter in the universe is a reversible adiabatic process and in any reversible
adiabatic process the entropy remains constant,
it increases only in irreversible processes \citep{Saha1950}. Consequently,
the contention that entropy of the universe would build up from oscillation to 
oscillation in the oscillatory model of the universe is not at all correct; oscillatory
model of the universe would not violate the second law of the thermodynamics.

\section*{Acknowledgements}
The author thanks Professor S. K. Pandey, Vice-Chancellor and the Co-ordinator of the 
University Grants Commission's Inter-University Centre for Astronomy and 
Astrophysics (IUCAA) Reference Centre, Pt. Ravishankar Shukla University, Raipur for
 making available the facilities of the centre to him. He also thanks Laxmikant Chaware
and Arun Kumar Diwaker for typing the manuscript.

 \bibliographystyle{mn2e}  
 \bibliography{RKTref}                

\begin{thebibliography}{68}
\expandafter\ifx\csname natexlab\endcsname\relax\def\natexlab#1{#1}\fi

\bibitem[{{Adams} {et~al.}(2005){Adams}, {Aggarwal}, \& {Ahammed}}]{Adams2005}
{Adams} J., {Aggarwal} M.~M., {Ahammed} Z., 2005, Nucl. Phys., A757, 102

\bibitem[{{Adcox} {et~al.}(2005){Adcox}, {Adler}, \& {Afanasiev}}]{Adcox2005}
{Adcox} K., {Adler} S.~S., {Afanasiev} S., 2005, Nucl. Phys., A757, 184

\bibitem[{{Alabiso} \& {Schierholz}(1976)}]{Alabiso1976}
{Alabiso} C., {Schierholz} G., 1976, Nuclear Physics B, 110, 81

\bibitem[{{Albrecht} \& {Steinhardt}(1982)}]{Albrecht1982}
{Albrecht} A., {Steinhardt} P.~J., 1982, Physical Review Letters, 48, 1220

\bibitem[{{Allen}(1973)}]{Allen1973}
{Allen} C.~W., 1973, {Astrophysical quantities}. The Athlone Press, London, p.
  293

\bibitem[{{Alpher}(1948)}]{Alpher1948b}
{Alpher} R.~A., 1948, Physical Review, 74, 1577

\bibitem[{{Alpher} {et~al.}(1948){Alpher}, {Bethe}, \& {Gamow}}]{Alpher1948a}
{Alpher} R.~A., {Bethe} H.~A., {Gamow} G., 1948, Physical Review, 73, 803

\bibitem[{{Alpher} \& {Herman}(1948)}]{Alpher1948}
{Alpher} R.~A., {Herman} R.~C., 1948, Physical Review, 74, 1737

\bibitem[{{Alpher} \& {Herman}(1950)}]{Alpher1950}
---, 1950, Reviews of Modern Physics, 22, 153

\bibitem[{{Aronson} \& {Ludlam}(2005)}]{Aronson2005}
{Aronson} S., {Ludlam} T., 2005,
  http://www.bnl.gov/npp/docs/hunting

\bibitem[{{Arsene} {et~al.}(2005){Arsene}, {Bearden}, \& {Beavis}}]{Arsene2005}
{Arsene} I., {Bearden} I.~G., {Beavis} D., 2005, Nuclear Physics A, 757, 1

\bibitem[{{Ashtekar} {et~al.}(1998){Ashtekar}, {Baez}, {Corichi}, \&
  {Krasnov}}]{Ashtekar1998}
{Ashtekar} A., {Baez} J., {Corichi} A., {Krasnov} K., 1998, Physical Review
  Letters, 80, 904

\bibitem[{{Ashtekar} {et~al.}(2001){Ashtekar}, {Baez}, \&
  {Krasnov}}]{Ashtekar2001}
{Ashtekar} A., {Baez} J.~C., {Krasnov} K., 2001, Adv. Theor. Math. Phys., 4, 1

\bibitem[{{Ashtekar} {et~al.}(2003){Ashtekar}, {Bojowald}, \&
  {Lewandowski}}]{Ashtekar2003}
{Ashtekar} A., {Bojowald} M., {Lewandowski} J., 2003, Adv. Theor. Math. Phys.,
  7, 233

\bibitem[{{Ashtekar} {et~al.}(2007){Ashtekar}, {Pawlowski}, {Singh}, \&
  {Vandersloot}}]{Ashtekar2007}
{Ashtekar} A., {Pawlowski} T., {Singh} P., {Vandersloot} K., 2007, \prd, 75,
  024035

\bibitem[{{Audouze}(1987a)}]{Audouze1987a}
{Audouze} J., 1987a, in IAU Symposium, Vol. 117, Dark matter in the universe,
  {J.~Kormendy \& G.~R.~Knapp}, ed., pp. 499--521

\bibitem[{{Audouze}(1987b)}]{Audouze1987b}
---, 1987b, in IAU Symposium, Vol. 124, Observational Cosmology, {A.~Hewitt,
  G.~Burbidge, \& L.~Z.~Fang}, ed., pp. 89--115

\bibitem[{{Back} {et~al.}(2005){Back}, {Baker}, \& {Ballintijn}}]{Back2005}
{Back} B.~B., {Baker} M.~D., {Ballintijn} M., 2005, Nucl. Phys. A, 757, 28

\bibitem[{{Balantekin}(2006)}]{Balantekin2006}
{Balantekin} A.~B., 2006, in Journal of Physics G: Nuclear and Particle Physics
  Review of Particle Physics, {Balantekin} A.~B., ed., Vol.~33, p.~36

\bibitem[{{Barber} {et~al.}(1979){Barber}, {Becker}, \& {Benda}}]{Barber1979}
{Barber} D.~P., {Becker} U., {Benda} H., 1979, Physical Review Letters, 43,
  1915

\bibitem[{{Baum} \& {Frampton}(2007{\natexlab{a}})}]{Baum2007a}
{Baum} L., {Frampton} P.~H., 2007{\natexlab{a}}, arXiv:hep-th/0703162

\bibitem[{{Baum} \& {Frampton}(2007{\natexlab{b}})}]{Baum2007b}
---, 2007{\natexlab{b}}, Physical Review Letters, 98, 071301

\bibitem[{{Beem}(1976)}]{Beem1976}
{Beem} J.~K., 1976, Comm. Math. Phys., 49, 179

\bibitem[{{Boesgaard} \& {Steigman}(1985)}]{Boesgaard1985}
{Boesgaard} A.~M., {Steigman} G., 1985, \araa, 23, 319

\bibitem[{{Bojowald}(2001)}]{Bojowald2001}
{Bojowald} M., 2001, Physical Review Letters, 86, 5227

\bibitem[{{Bojowald}(2005)}]{Bojowald2005}
---, 2005, Journal of Physics Conference Series, 24, 77

\bibitem[{{Bojowald}(2007)}]{Bojowald2007}
---, 2007, Nature Physics, 3, 523

\bibitem[{{Chaichian} {et~al.}(1981){Chaichian}, {Demichev}, \&
  {Nelipa}}]{Chaichian1981}
{Chaichian} M., {Demichev} A.~P., {Nelipa} N.~F., 1981, Physics Letters B, 102,
  43

\bibitem[{{Dicke} \& {Peebles}(1979)}]{Dicke1979}
{Dicke} R.~H., {Peebles} P.~J.~E., 1979, in General Relativity: An Einstein
  centenary survey, {S.~W.~Hawking \& W.~Israel}, ed., pp. 504--517

\bibitem[{Frampton(2006)}]{Frampton2006}
Frampton P.~H., 2006, arXiv:astro-ph/0612243

\bibitem[{{Friedmann}(1922)}]{Friedmann1922}
{Friedmann} A., 1922, Zeitschrift fur Physik, 10, 377

\bibitem[{{Friedmann}(1924)}]{Friedmann1924}
---, 1924, Zeitschrift fur Physik, 21, 326

\bibitem[{{Gamow}(1946)}]{Gamow1946}
{Gamow} G., 1946, Physical Review, 70, 572

\bibitem[{{Gamow}(1948)}]{Gamow1948}
---, 1948, Physical Review, 74, 505

\bibitem[{{Geroch}(1966)}]{Geroch1966}
{Geroch} R.~P., 1966, Physical Review Letters, 17, 445

\bibitem[{{Geroch}(1968)}]{Geroch1968}
---, 1968, Annals of Physics, 48, 526

\bibitem[{Greenberg(1964)}]{Greenberg1964}
Greenberg O.~W., 1964, Phys. Rev. Lett., 13, 598

\bibitem[{{Gross} \& {Wilczek}(1973{\natexlab{a}})}]{Gross1973a}
{Gross} D.~J., {Wilczek} F., 1973{\natexlab{a}}, \prd, 8, 3633

\bibitem[{{Gross} \& {Wilczek}(1973{\natexlab{b}})}]{Gross1973b}
---, 1973{\natexlab{b}}, Physical Review Letters, 30, 1343

\bibitem[{{Guth}(1981)}]{Guth1981}
{Guth} A.~H., 1981, \prd, 23, 347

\bibitem[{{Hawking} \& {Penrose}(1970)}]{Hawking1970}
{Hawking} S.~W., {Penrose} R., 1970, Royal Society of London Proceedings Series
  A, 314, 529

\bibitem[{{Heinz} \& {Jacob}(2000)}]{Heinz2000}
{Heinz} U., {Jacob} M., 2000, arXiv:nucl-th/0002042v1

\bibitem[{{Hofstadter} \& {Mc Allister}(1955)}]{Hofstadter1955}
{Hofstadter} R., {Mc Allister} R.~W., 1955, Physical Review, 98, 217

\bibitem[{{Infeld} \& {Schild}(1945)}]{Infeld1945}
{Infeld} L., {Schild} A., 1945, Physical Review, 68, 250

\bibitem[{{Israelit} \& {Rosen}(1989)}]{Israelit1989}
{Israelit} M., {Rosen} N., 1989, \apj, 342, 627

\bibitem[{{Kembhavi}(1978)}]{Kembhavi1978}
{Kembhavi} A.~K., 1978, \mnras, 185, 807

\bibitem[{{Linde}(1982)}]{Linde1982}
{Linde} A.~D., 1982, Physics Letters B, 108, 389

\bibitem[{{Linde}(1983)}]{Linde1983}
---, 1983, Physics Letters B, 129, 177

\bibitem[{{Nambu}(1966)}]{Nambu1966}
{Nambu} Y., 1966, in A. de Shalit (ed.), Preludes in Theoretical Physics,
  North-Holland, Amsterdam

\bibitem[{{Narlikar} \& {Padmanabhan}(1986)}]{Narlikar1986}
{Narlikar} J.~V., {Padmanabhan} T., 1986, Gravity, gauge theories and quantum
  cosmology. D. Reidel Publ. Co., Dordrecht, Holland

\bibitem[{{Olive} {et~al.}(1981){Olive}, {Schramm}, {Turner}, {Yang}, \&
  {Steigman}}]{Olive1981}
{Olive} K.~A., {Schramm} D.~N., {Turner} M.~S., {Yang} J., {Steigman} G., 1981,
  \apj, 246, 557

\bibitem[{{Penrose}(1965)}]{Penrose1965}
{Penrose} R., 1965, Physical Review Letters, 14, 57

\bibitem[{{Penzias} \& {Wilson}(1965)}]{Penzias1965}
{Penzias} A.~A., {Wilson} R.~W., 1965, \apj, 142, 419

\bibitem[{{Politzer}(1973)}]{Politzer1973}
{Politzer} H.~D., 1973, Physical Review Letters, 30, 1346

\bibitem[{{Rindler}(1956)}]{Rindler1956}
{Rindler} W., 1956, \mnras, 116, 662

\bibitem[{{Saha} \& {Srivastava}(1950)}]{Saha1950}
{Saha} M.~N., {Srivastava} B.~N., 1950, A Treatise on Heat. The Indian Press
  Ltd., Allahabad, p. 316

\bibitem[{{Starobinsky}(1980)}]{Starobinsky1980}
{Starobinsky} A.~A., 1980, Physics Letters B, 91, 99

\bibitem[{{Steigman}(1976)}]{Steigman1976}
{Steigman} G., 1976, \araa, 14, 339

\bibitem[{{Steigman}(1979)}]{Steigman1979}
---, 1979, Annual Review of Nuclear and Particle Science, 29, 313

\bibitem[{Steinhardt \& Turok(2001{\natexlab{a}})}]{Steinhardt2001a}
Steinhardt P.~J., Turok N., 2001{\natexlab{a}}, arXiv:hep-th/0111098

\bibitem[{Steinhardt \& Turok(2001{\natexlab{b}})}]{Steinhardt2001b}
---, 2001{\natexlab{b}}, arXiv:hep-th/0111030

\bibitem[{Steinhardt \& Turok(2004)}]{Steinhardt2004}
---, 2004, arXiv:astro-ph/0404480

\bibitem[{{Thakur}(1992)}]{Thakur1992}
{Thakur} R.~K., 1992, \apss, 190, 281

\bibitem[{{Thakur}(1995)}]{Thakur1995}
---, 1995, \ssr, 73, 273

\bibitem[{{Thakur}(2009)}]{Thakur2009}
---, 2009, arXiv:0901.1956v1

\bibitem[{{Weinberg}(1972{\natexlab{a}})}]{weinberg1972}
{Weinberg} S., 1972{\natexlab{a}}, {Gravitation and Cosmology: Principles and
  Applications of the General Theory of Relativity}. Wiley, New York, p. 472

\bibitem[{{Weinberg}(1972{\natexlab{b}})}]{Weinberg1972a}
---, 1972{\natexlab{b}}, {Ibid}. p. 147, see also the preface of the book,
  epecially the third and the fourth paragraphs.

\bibitem[{{Yang} {et~al.}(1984){Yang}, {Turner}, {Schramm}, {Steigman}, \&
  {Olive}}]{Yang1984}
{Yang} J., {Turner} M.~S., {Schramm} D.~N., {Steigman} G., {Olive} K.~A., 1984,
  \apj, 281, 493

\end{thebibliography}


%

\end{document}